\documentclass[conference]{IEEEtran}
\IEEEoverridecommandlockouts
\usepackage{cite}
\usepackage{amsmath,amssymb,amsfonts}
\usepackage{algorithmic}
\usepackage{graphicx}
\usepackage{textcomp}
\usepackage{xcolor}
\def\BibTeX{{\rm B\kern-.05em{\sc i\kern-.025em b}\kern-.08em
    T\kern-.1667em\lower.7ex\hbox{E}\kern-.125emX}}
    
\usepackage{acronym}
\newacro{sme}[SME]{Small and Medium-sized Enterprise}
\newacro{it}[IT]{Information Technology}
\newacro{ot}[OT]{Operation Technology}
\newacro{cps}[CPS]{Cyber-Physical System}
\newacro{cpps}[CPPS]{Cyber-Physical Production System}
\newacro{ids}[IDS]{Intrusion Detection System}
\newacro{svm}[\textit{SVM}]{\textit{Support Vector Machine}}
\newacro{wsn}[WSN]{Wireless Sensor Network}
\newacro{darpa}[DARPA]{Defense Advanced Research Projects Agency}
\newacro{kdd}[KDD]{Knowledge Discovery in Databases}
\newacro{scada}[SCADA]{Supervisory Control And Data Acquisition}
\newacro{dpi}[DPI]{Deep Packet Inspection}
\newacro{dmz}[DMZ]{De-Militarized Zone}

\usepackage{booktabs}

\usepackage{multirow}
    
\begin{document}

\title{Modern Problems Require Modern Solutions:\\Hybrid Concepts for Industrial Intrusion Detection
\thanks{This is a preprint of a publication published at the 24. ITG Fachtagung Mobilkommunikation.
Please cite as: S. D. Duque Anton, M. Strufe, H. D. Schotten, ``Modern Problems Require Modern Solutions: Hybrid Concepts for Industrial Intrusion Detection,'' in \textit{24. ITG Fachtagung Mobilkommunikation - Technologien und Anwendungen (MKT-19)}.  Informationstechnische Gesellschaft im VDE (ITG), VDE Verlag GmbH, 2019, pp. 105-109.}
}

\author{\IEEEauthorblockN{1\textsuperscript{st} Simon D. Duque Anton}
\IEEEauthorblockA{\textit{Intelligent Networks Research Group} \\
\textit{German Research Center for AI}\\
67663 Kaiserslautern, Germany\\
Simon.Duque\_Anton@dfki.de}
\and
\IEEEauthorblockN{2\textsuperscript{nd} Mathias Strufe}
\IEEEauthorblockA{\textit{Intelligent Networks Research Group} \\
\textit{German Research Center for AI}\\
67663 Kaiserslautern, Germany\\
Mathias.Strufe@dfki.de}
\and
\IEEEauthorblockN{3\textsuperscript{rd} Hans Dieter Schotten}
\IEEEauthorblockA{\textit{Intelligent Networks Research Group} \\
\textit{German Research Center for AI}\\
67663 Kaiserslautern, Germany\\
Hans\_Dieter.Schotten@dfki.de}
}

\maketitle

\begin{abstract}
The concept of Industry 4.0 brings a disruption into the processing industry.
It is characterised by a high degree of intercommunication, embedded computation,
resulting in a decentralised and distributed handling of data.
Additionally,
cloud-storage and Software-as-a-Service (SaaS) approaches enhance a centralised storage and handling of data.
This often takes place in third-party networks.
Furthermore,
Industry 4.0 is driven by novel business cases.
Lot sizes of one,
customer individual production,
observation of process state and progress in real-time and remote maintenance,
just to name a few.
All of these new business cases make use of the novel technologies.
However,
cyber security has not been an issue in industry.
Industrial networks have been considered physically separated from public networks.
Additionally,
the high level of uniqueness of any industrial network was said to prevent attackers from exploiting flaws.
Those assumptions are inherently broken by the concept of Industry 4.0.
As a result,
an abundance of attack vectors is created.
In the past,
attackers have used those attack vectors in spectacular fashions.
Especially Small and Medium-sized Enterprises (SMEs) in Germany struggle to adapt to these challenges.
Reasons are the cost required for technical solutions and security professionals.
In order to enable SMEs to cope with the growing threat in the cyberspace,
the research project IUNO Insec aims at providing and improving security solutions that can be used without specialised security knowledge.
The project IUNO Insec is briefly introduced in this work.
Furthermore,
contributions in the field of intrusion detection,
especially machine learning-based solutions,
for industrial environments provided by the authors are presented and set into context.
\end{abstract}


\section{Introduction}
In the past two decades,
attacks on industrial systems have increased in a spectacular manner~\cite{Duque_Anton.2017a}.
Malicious actors,
supposedly state-sponsored and criminal groups,
have taken an interest in industrial organisations.
Attacks on critical infrastructures,
such as the power grid in the Ukraine in December 2015~\cite{Cherepanov.2017},
have a severe impact,
potentially matching a political agenda.
Attributing these attacks,
however,
is a non-trivial task~\cite{Fraunholz.2017d, Fraunholz.2017f}.
Espionage and sabotage to gain financial or technical benefits and hinder a competitor are scenarios that need to be taken into consideration as well.
In Germany,
hundreds of so-called hidden champions,
enterprises unknown by the customer but leading in their area,
exist~\cite{Bayley.2017}.
Despite their importance for economy,
they are often \acp{sme}.
Due to their size,
\ac{it} security is commonly not a priority.
Investments in \ac{it} security are expensive in terms of personnel and resources but do not provide measurable revenue.
Additionally,
increases in \ac{it} security measures are commonly perceived as a hindrance to productivity.
Non-technical requirements on \ac{it} security measures for application in \acp{sme} include ease of use,
scalability and usability by non-experts in the security field.
The publicly funded research project IUNO~\cite{IUNO.} aimed at providing such measures in order for industrial enterprises to integrate them into their production environments.
After it was finshed,
the research project IUNO Insec~\cite{Insec.} was established as a succeeding research project.
The scope is the improvement and further development of the tools implemented in IUNO.
In this work,
the research project IUNO Insec is presented.
Furthermore, different aspects of intrusion detection suitable for industrial environments are presented.
Apart from the \ac{it} networks,
industrial environments contain control networks or \ac{ot}.
Legacy protocols and security not being a design target make \ac{ot} networks vulnerable to attackers.
The usage of \acp{cps} and \acp{cpps} in \ac{ot} networks enables attackers to have an effect on the physical world by means of the digital world. \\ \par
The remainder of this work is structured as follows.
In Section~\ref{sec:sota},
a brief overview of the state of the art is provided.
The research project IUNO Insec and its goals are introduced in Section~\ref{sec:insec}.
An exemplary summary of intrusion detection methods developed for industrial use is presented in Section~\ref{sec:ids}.
Potentially beneficial combinations of those methods are discussed in Section~\ref{sec:combining}.
This work is concluded in Section~\ref{sec:conc}.

\section{State of the Art}
\label{sec:sota}
Due to the relevance of critical infrastructures on supply chains and nations,
the effects of incidents on production and revenue and the dangers of digital attacks on phyiscal systems,
information security has gained importance in the industrial community.
Therefore,
an industry focusing on industrial intrusion detection solutions has evolved.
Additionally,
the research community has taken an interest in solutions for industrial \acp{ids}.
After the first work of \textit{Denning} in 1987~\cite{Denning.1987},
intrusion detection has become established as a field of research and application.
A brief overview of intrusion detection approaches is provided in Table~\ref{tab:sota}.
\begin{table}[h!]
\renewcommand{\arraystretch}{1.3}
\caption{Applications of Anomaly Detection by the Individual Works}
\label{tab:sota}
\centering
\scriptsize
\begin{tabular}{l l}
\toprule
\textbf{Subject Covered} & \textbf{Research Work} \\
Surveys \& Taxonomies & \cite{Zhu.2011, Garcia-Teodoro.2009, Lee.1998} \\
Wireless Networks & \cite{Shin.2010, Zhang.2003} \\
Industrial Networks & \cite{Caselli.2015, Morris.2012, Shin.2010}\\
\ac{it} Networks & \cite{Northcutt.2002, Mukherjee.1994} \\
Machine Learning-based & \cite{Mukkamala.2002, Ryan.1998} \\
\bottomrule
\end{tabular}
\end{table}
\textit{Zhu} discusses attacks on \ac{scada} systems~\cite{Zhu.2011}.
A summary of existing systems as well as challenges in anomaly-based network intrusion detection is provided by \textit{Garcia-Teodoro et al.}~\cite{Garcia-Teodoro.2009}.
\textit{Lee and Stolfo} provide an overview of data mining approaches for intrusion detection~\cite{Lee.1998}.
\textit{Zhang et al.} analyse techniques to detect intrusions in mobile networks~\cite{Zhang.2003}.
Detecting intrusions in mobile ad-hoc networks is discussed by \textit{Zhang and Wenke} as well~\cite{Zhang.2000}.
\textit{Shin et al.} discuss intrusion detection for industrial wireless networks~\cite{Shin.2010}.
This is motivatied by the increasing relevance of wireless sensors,
organised in so-called \acp{wsn}.
Industrial networks often contain fixed sequences of operations.
These can be exploited to detect deviations,
as performed by ~\textit{Caselli et al.}~\cite{Caselli.2015}.
Many industrial networks contain legacy protocols that have to be integrated into working intrusion detection solutions.
\textit{Morris et al.} present an intrusion detection system to fit \textit{Modbus} communication~\cite{Morris.2012}.
Detection of intrusions in \ac{it} networks is discussed by \textit{Northcutt and Novak}~\cite{Northcutt.2002}.
\textit{Mukherjee et al.} present work on detecting intrusions in networks~\cite{Mukherjee.1994}.
\textit{Mukkamala et al.} employ neural networks and \acp{svm} to detect intrusions in the infamous \ac{darpa} \ac{kdd} cup '99~\cite{Mukkamala.2002}.
\textit{Ryan et al.} employ neural networks as well to detect intrusions~\cite{Ryan.1998}.

\section{Enter IUNO Insec}
\label{sec:insec}
\begin{figure*}[h!]
\centering
\includegraphics[width=0.98\textwidth]{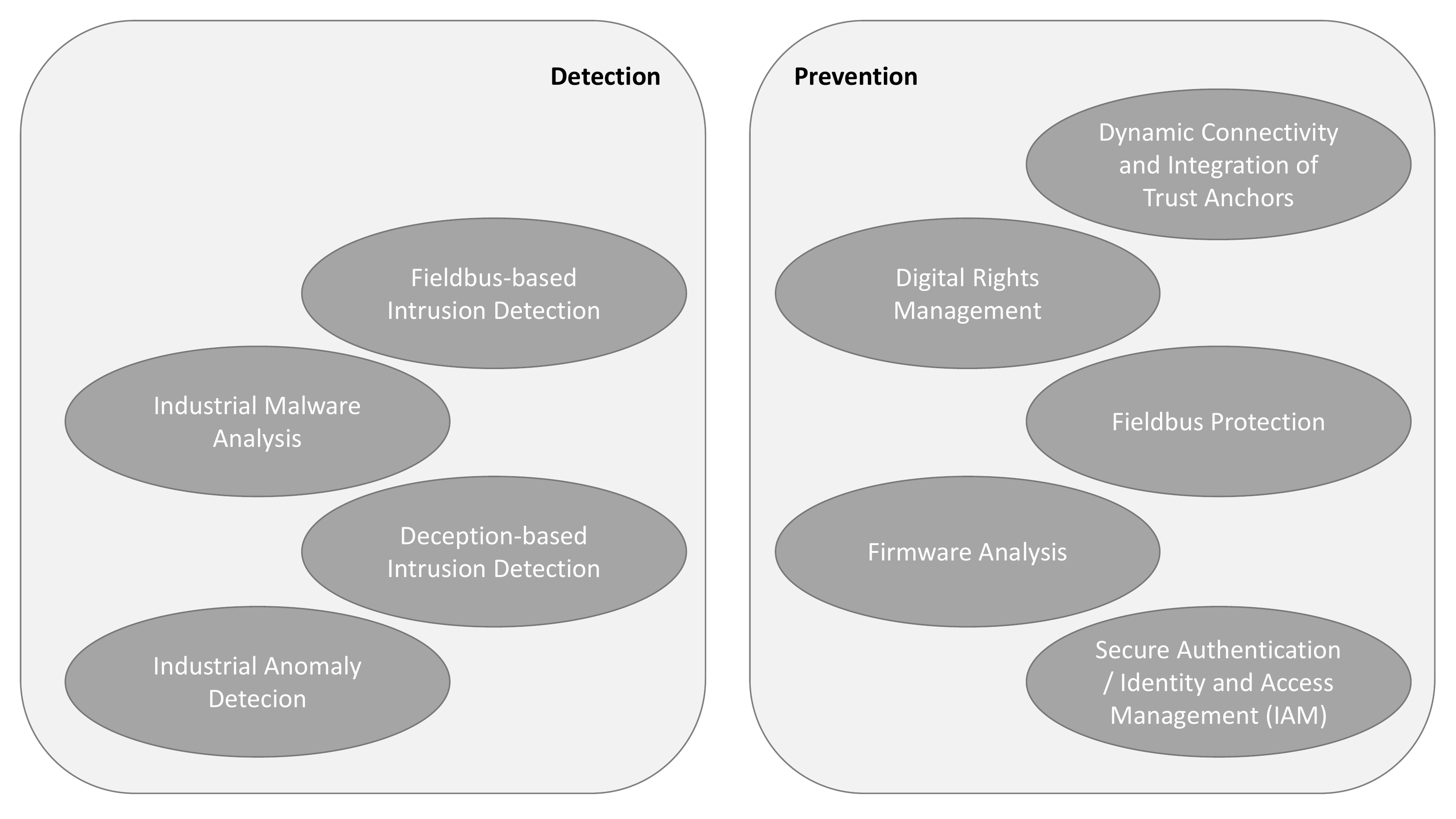}
\caption{Methods and Tools Addressed by IUNO Insec}
\label{fig:topics}
\end{figure*}

IUNO Insec~\cite{Insec.} is the succeeding research project of the national reference project for \ac{it} security in Industry 4.0 (IUNO)~\cite{IUNO.}.
In IUNO,
four application uses cases were addressed:
\begin{itemize}
\item Customer-individual production
\item Technology data market place
\item Remote maintenance
\item Visual Security Control
\end{itemize}
These application use cases were derived in coordination with the industrial partners,
ensuring their relevance.
Solutions were developed, 
with different degrees of technology readiness.
They were then collected in the toolbox used to provide the solutions to interested third parties.
As the goal of IUNO was to provide means to secure their digital assets to \acp{sme},
the toolbox was published.
However,
many solutions were introduced by academic partners,
lacking technology readiness to be used as they were.
In order to increase technology readiness,
refine the tools and adjust them to different application use cases,
the research project IUNO Insec was established.
The main areas of research and development in IUNO Insec are shown in Figure~\ref{fig:topics}.
The two major security controls addressed by IUNO Insec are detection and prevention of malicious activities,
meaning the goal is to either hinder an attacker from performing malicious acts or detecting the process of an attacker doing so.
Other security controls,
such as recovery,
prevention or correction are not in the scope of IUNO Insec.
Automated tools that actively engage in activities in an industrial networks always contain the danger of negatively influencing production.
This could amount to enormous costs and is thus undesirable in industrial use cases,
as opposed to home and office environments where short disruptions of the productive system are tolerable.
Since industrial environments are commonly highly application specific,
security solutions need to be capable of abstracting and transferring functionality.
Furthermore,
they need to be usable by non-experts while still meeting functional and non-functional requirements of the users.
User-centric workshops are organised to ensure these features.

\section{Industrial Intrusion Detection Approaches}
\label{sec:ids}
In the course of research project IUNO,
the German Research Center for Artificial Intelligence (DFKI) has evaluated and implemented several solutions to detecting anomalies in industrial networks and processes.
Anomalies are events in a set of events that deviate with respect to certain features.
Even though the correspondence of an anomaly to an intrusion is not always given and non-trivial to evaluate,
anomalies are events a human operator should take a look at.
In this work,
analyses of packet-based intrusion detection are presented,
as well as time series-based anomaly detection in \ac{ot} network traffic and process data.

\subsection{Packet-based Detection}
Detecting intrusions based on features of singular network packets is a well-researched method~\cite{Roesch.1999}.
Each network packet contains metainformation,
such as source and destination,
size,
ports and many more.
Furthermore,
each packet contains a payload that,
given it is not encrypted,
can be used to determine malicious intents.
A special kind of packet-based intrusion detection is \ac{dpi} where a sequence of packets is considered.
This is relevant as many attacks,
e.g. the ACK-scan,
rely on disruptions of the sequences while not providing packets that are anomalous in themselves. \\ \par 
In this work,
data sets provided by \textit{Lemay and Fernandez} are evaluated with respect to packets~\cite{Lemay.2016}.
They provided data sets monitored in an emulated environment,
simulation power circuit breakers with \textit{Modbus}-based communication.
In some data sets, 
operations of a human were simulated.
After monitoring,
different kinds of attacks were introduced to the data sets.
However,
all attacks are TCP/IP-based,
none exploits \textit{Modbus}-specific vulnerabilities as presented by \textit{Morris et al.}~\cite{Morris.2013}.
The data sets used in this work and their corresponding name in the work of \textit{Lemay and Fernandez} are listed in Table~\ref{tab:data_sets_lemay}.
\begin{table}[h!]
\setlength{\tabcolsep}{4pt}
\renewcommand{\arraystretch}{1.3}
\caption{Data Sets Used as Presented by \textit{Lemay and Fernandez}}
\label{tab:data_sets_lemay}
\centering
\scriptsize
\begin{tabular}{l l r r}
\toprule
\textbf{ID} & \textbf{Name} & \textbf{Len. (s)} & \textbf{Attacks}  \\
\textit{DS1} & Moving\_two\_files\_Modbus\_6RTU & 190 & 4 \\
\textit{DS2} & Send\_a\_fake\_command\_Modbus\_6RTU\_with\_operate & 670 & 1 \\
\textit{DS3} & CnC\_uploading\_exe\_modbus\_6RTU\_with\_operate & 70 & 2 \\
\bottomrule
\end{tabular}
\end{table}
In a previous work~\cite{Duque_Anton.2018b},
these data sets have been evaluated with four different anomaly detection algorithms:
\begin{itemize}
\item \textit{Random Forest}
\item \ac{svm}
\item \textit{k-nearest Neighbour}
\item \textit{k Means Clustering}
\end{itemize}
In order to evaluate the performance,
the accuracy as shown in (\ref{eq:accuracy}) as well as the f1-score (\ref{eq:f-measure}) that is based on precision (\ref{eq:precision}) and recall (\ref{eq:recall}) are used.
\begin{equation}
\label{eq:accuracy}
accuracy = \dfrac{t_{p} + t_{n}}{t_{p}+f_{p} + t_{n}+f_{n}}
\end{equation}
\begin{equation}
\label{eq:f-measure}
F_{1} = 2 \cdot \dfrac{precision \cdot recall}{precision + recall}
\end{equation}
\begin{equation}
\label{eq:precision}
precision = \dfrac{t_{p}}{t_{p}+f_{p}}
\end{equation}
\begin{equation}
\label{eq:recall}
recall = \dfrac{t_{p}}{t_{p}+f_{n}}
\end{equation}
$t_{p}$ denotes the positive events that are classified correctly,
$t_{n}$ the negative ones that are classified correctly.
$f_{p}$ and $f_{n}$ denote the positive and negative events respectively that are misclassified.
The results are shown in Table~\ref{tab:packet_res}.
\begin{table}[h!]
\renewcommand{\arraystretch}{1.3}
\caption{Results of Packet-based Anomaly Detection}
\label{tab:packet_res}
\centering
\scriptsize
\begin{tabular}{l l r r r}
\toprule
\textbf{Algorithm} & \textbf{Metric} & \textbf{\textit{DS1}} & \textbf{\textit{DS2}} & \textbf{\textit{DS3}}  \\
\cmidrule{2-5}
\multirow{2}{*}{\textit{Random Forest}} & Acc. & 1.0 & 0.99970 & 0.99997 \\
& F1 & 1.0 & 0.99985 & 0.99999 \\
\cmidrule{2-5}
\multirow{2}{*}{\ac{svm}} & Acc. & 1.0 & 1.0 & 0.99994 \\
& F1 & 1.0 & 1.0 & 0.99997 \\
\cmidrule{2-5}
\multirow{2}{*}{\textit{k-nearest Neighbour}} & Acc. & 0.99710 & 0.99912 & 0.99941 \\
& F1 & 0.99853 & 0.99956 & 0.99971 \\
\cmidrule{2-5}
\multirow{2}{*}{\textit{k Means Clustering}} & Acc. & 0.98102 & 0.55624 & 0.63362 \\
& F1 & 0.99038 & 0.71485 & 0.77573 \\
\bottomrule
\end{tabular}
\end{table}
It can be seen that \ac{svm} and \textit{Random Forest} perform very well with near perfect scores,
while \textit{k-nearest Neighbour} performs well in certain areas and \textit{k Means Clustering} does not perform satisfactorily at all,
allowing too many false positives.

\subsection{Time Series-based Detection in Network Traffic}
\begin{figure}[h!]
\centering
\includegraphics[width=0.48\textwidth]{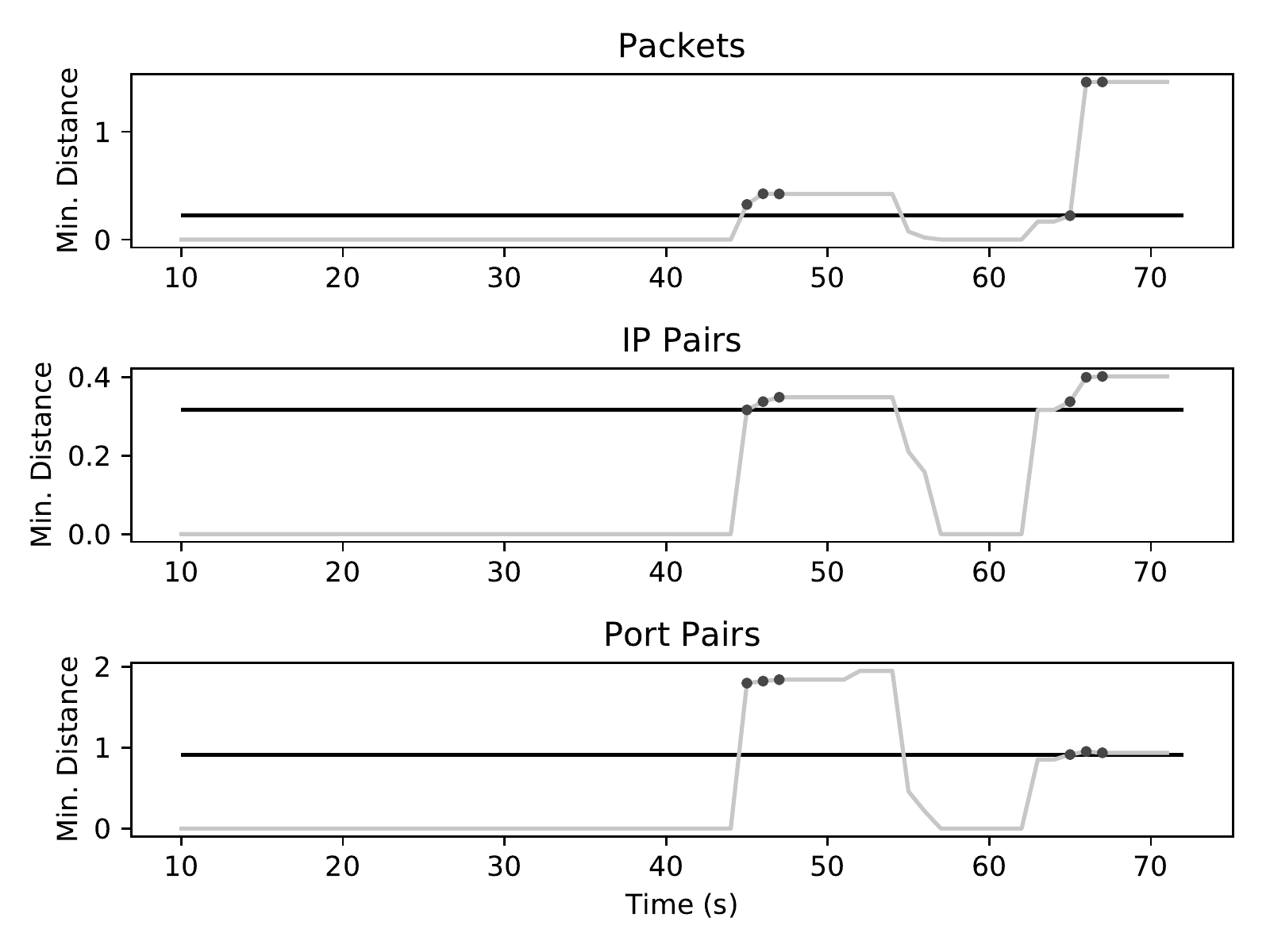}
\caption{Time Series-based Intrusion Detection in Industrial Network Traffic}
\label{fig:ts_nw}
\end{figure}

Industrial network communication is expected to be highly periodic.
Processes produce repeating patterns in communication,
the number and structure of entities communicating is expected to stay constant.
In order to detect deviations in the time domain,
time series anomaly detection was applied in a previous work~\cite{Duque_Anton.2018c}.
As an algorithm,
\textit{Matrix Profiles} were used,
as presented by \textit{Yeh et al.}~\cite{Yeh.2016a}.
The concept of \textit{Matrix Profiles} is the calculation of distances between sequences of length $m$.
Any sequence of length $m$ is compared to any other sequence of length $m$.
Then the minimal distance is derived.
Originally an algorithm for motif discovery,
\textit{Matrix Profiles} can detect outliers if the minimal distance is high.
This indicates that a sequence was singular,
being a hint on anomalous behaviour.
The minimal distances of packet number,
port and IP pairs of \textit{DS3} as defined in Table~\ref{tab:data_sets_lemay} is shown in Figure~\ref{fig:ts_nw}.
It shows that the \textit{Matrix Profiles} can be used to detect any attack with the black line showing a minimal threshold value needed to create no false negatives.
The length of increases in minimal distance can be explained with the window length $m$,
as any attack influences all $m$ following sequences as well.
However,
automatically detecting a threshold that performs well is a non-trivial challenge to be solved.

\subsection{Time Series-based Detection in Process Data}
\begin{figure}[h!]
\centering
\includegraphics[width=0.48\textwidth]{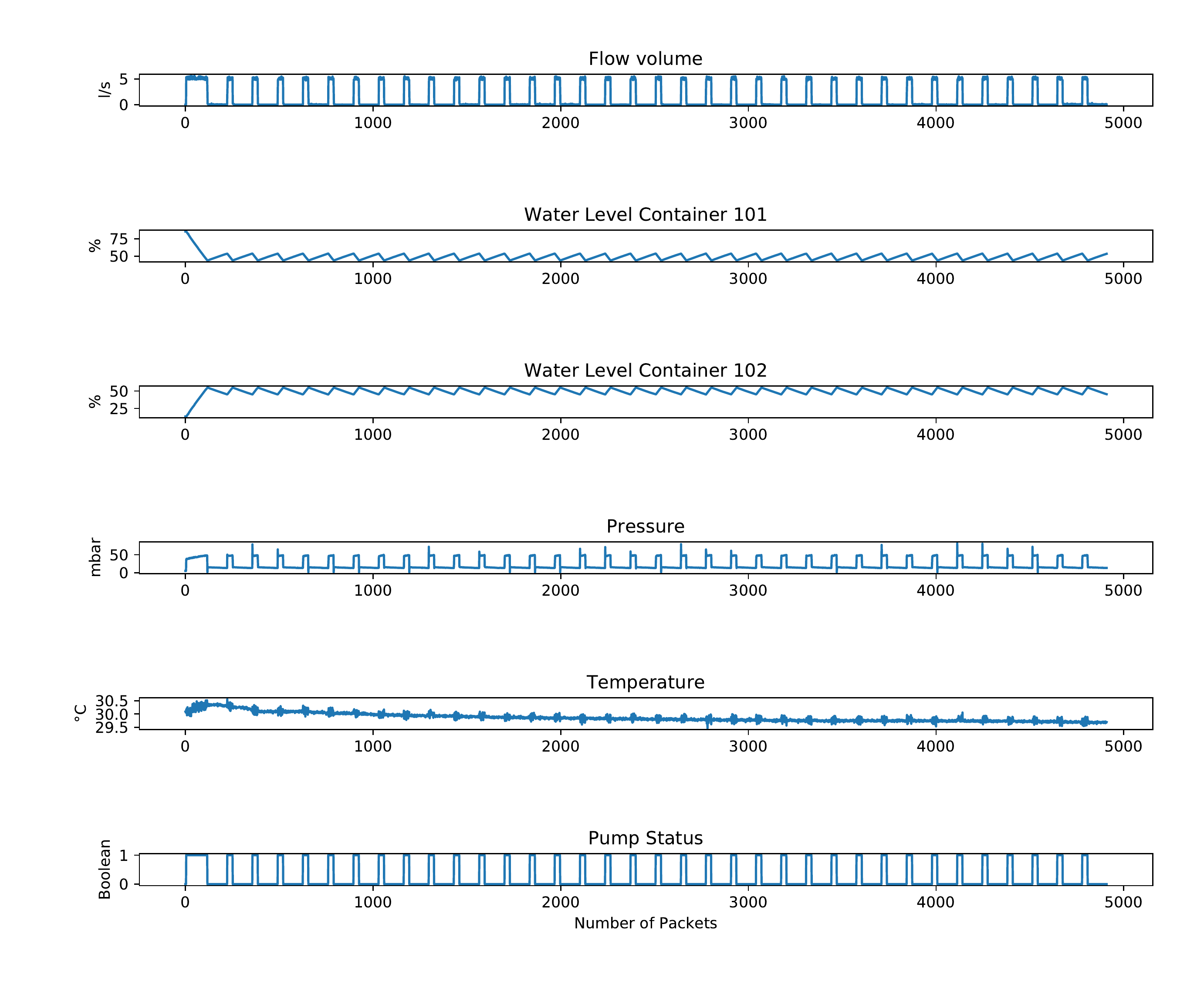}
\caption{Real-world Time Series of Process}
\label{fig:real_proc}
\end{figure}

Apart from the network traffic characteristics,
time series can be used to detect deviations in the process behaviour.
An exemplary process using batch processing has been created using real world hardware~\cite{Duque_Anton.2019a}.
It is shown in Figure~\ref{fig:real_proc}.
A pump is used to pump water from Container 102 to Container 101.
Due to natural reflow,
the level of Container 101 decreases over time.
A hysteresis value activates the pump again once too much water has reflown.
This process has been the base for a simulated environment,
introducing attacks to a simulation of the same process~\cite{Duque_Anton.2019c}.
The water flow as well as the water level of Container 102  are shown in c.
\begin{figure}[h!]
\centering
\includegraphics[width=0.48\textwidth]{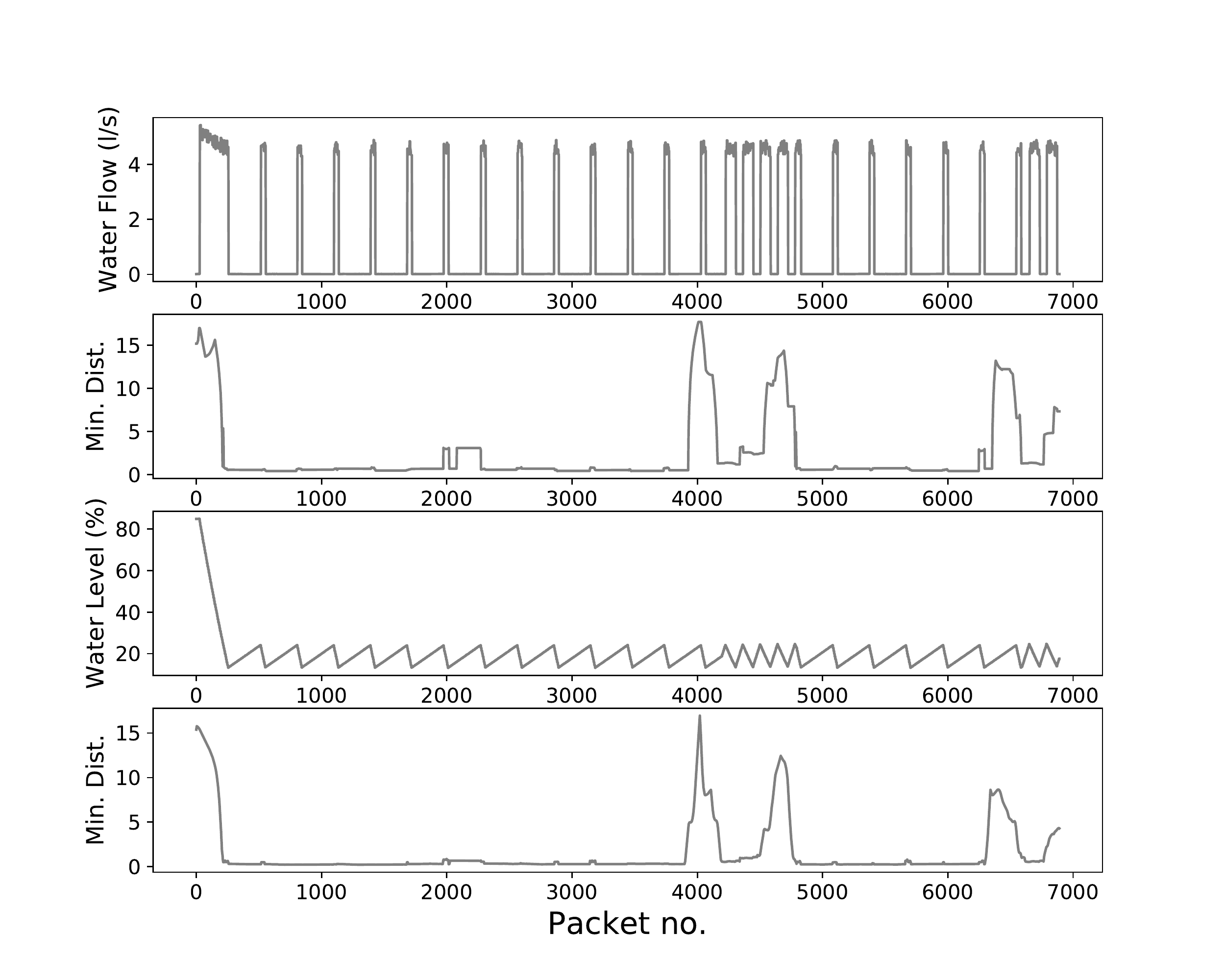}
\caption{Time Series-based Intrusion Detection in Industrial Process Data}
\label{fig:ts_proc}
\end{figure}
Normal operation is pictured until packet number 4000.
After that and until packet 4800,
the speed of reflow is doubled, 
halving the periods of pump activity and the water level refill.
The same behaviour is shown again starting at packet number 6500 to the end of the trace.
Minimal distances as calculated with \textit{Matrix Profiles} rises significantly during the attacks as clearly indicated in Figure~\ref{fig:ts_proc}.
As a result,
time series-based anomaly detection methods are capable of detecting deviations and anomalies in regular process behaviour that can be indicators of attacks.

\section{Combining the Pieces}
\label{sec:combining}
As discussed,
attacks on industrial networks have to undergo different stages~\cite{Duque_Anton.2019c}.
First,
the perimeter has to be breached,
commonly by phishing or other social engineering attacks.
This grants an attacker access to the \ac{it} network.
The \ac{it} network is commonly connected to \ac{ot} networks by \acp{dmz},
network segmentation and firewalls.
Moving laterally from \ac{it} to \ac{ot} networks is crucial for attackers in order to execute the intended activity.
If the perimeter has been breached without triggering counter measures,
the lateral movement is the next phase of an attack to be detected.
A context-based aggregation model has been introduced that allows for distributed collection of data in order to determine sources,
destinations and effects of attacks~\cite{Duque_Anton.2017b}.
Context information can provide valuable insight on attacks and aid in detecting anomalies~\cite{Duque_Anton.2017c}.

\section{Conclusion}
\label{sec:conc}
In this work,
the research project IUNO Insec was introduced.
It aims at providing much needed security solutions for German industrial enterprises,
especially \acp{sme} that cannot afford to build expertise themselves.
The approach of IUNO Insec includes the improvement and development of easy to use security modules.
Furthermore,
approaches to detect anomalies in an industrial context are discussed.
Industrial environments contain characteristic requirements for security solutions as well as unique properties with respect to topology and communication behaviour.
This motivates the use of certain technologies,
such as time series or packet-based analysis.
Additionally,
deceptive technologies,
such as feints, distraction and obfuscation~\cite{Fraunholz.2018b} or honeypots~\cite{Fraunholz.2017c} can provide an additional layer of security to provide German \acp{sme} the means to securely participate in the fourth industrial revolution.

\section*{Acknowledgment}
This work has been supported by the Federal Ministry of Education and Research of the Federal Republic of Germany (Foerderkennzeichen 16KIS0932, IUNO Insec).
The authors alone are responsible for the content of the paper.

\bibliographystyle{IEEEtran}
\bibliography{literature}

\end{document}